\documentclass[journal]{IEEEtran}
\ifCLASSINFOpdf
   \usepackage[pdftex]{graphicx}
   \DeclareGraphicsExtensions{.pdf,.jpeg,.png}
\else
\fi
\usepackage{color}

\usepackage{amsmath}
\usepackage{textcase}

\begin{document}
%
%
%
%


\title{Novel Rigorous Multimode Equivalent Network Formulation for Boxed Planar Circuits with Arbitrarily Shaped Metallizations}       

\author{Celia~G\'omez~Molina, Fernando~Quesada~Pereira, Alejandro~Alvarez~Melcon,	 Stephan Marini,  Miguel A. S\'anchez-Soriano, Vicente~E.~Boria, and~Marco~Guglielmi
	
	 \thanks{
	 		This work has been supported by the Spanish Government through the Ministerio de Educaci\'on, Cultura y Deporte with Ref. FPU15/02883, the Ministerio de Econom\'ia y Competitividad through the sub-projects 4, 2 and 1 of the coordinated project with Ref. TEC2016-75934-C4-R and the Fundaci\'on S\'eneca de la Regi\'on de Murcia with Ref. 19494/PI/14 and Ref. 20147/EE/17. This paper is an expanded version from the International Conference on Numerical Electromagnetic and Multiphysics Modeling and Optimization, Reykjavik,	Iceland, August 8-10}

      	\thanks{C. G\'omez Molina, F. D. Quesada Pereira and A.~\'Alvarez~Melc\'on are with the Department of Information Technologies~and~Communications, Universidad Polit\'ecnica de~Cartagena,~30202~Cartagena,~Spain (e-mail:  celia.gomez@upct.es; fernando.quesada@upct.es; alejandro.alvarez@upct.es).}
      	\thanks{S. Marini and  M. A. S\'anchez-Soriano are with the Department of Physics, Systems Eng. \& Signal Theory, Universidad de Alicante,  03690~Alicante,~Spain  (e-mail: smarini@ua.es). }      	
      	\thanks{V. E. Boria and M. Guglielmi are with the R\&D Institute on Telecommunication \& Multimedia Applications,  Universidad Polit\'ecnica de Valencia, 46022~Valencia,~Spain  (e-mail: vboria@dcom.upv.es; marco.guglielmi@iteam.upv.es). }}

\maketitle

\begin{abstract}
The Multimode Equivalent Network (MEN) formulation has been traditionally employed for the efficient and accurate analysis of waveguide devices. In this paper, we extend the use of the MEN to the analysis of zero thickness, multilayer planar circuits in a metallic enclosure. The formulation is developed for arbitrary shape metallic areas and includes both internal and external ports in the transverse plane. The BI-RME method is applied to analyze the arbitrary shaped metallizations and to compute the coupling integrals needed to solve the corresponding integral equations. On this basis, typical shielded microstrip circuits of complex geometries, are analyzed in the common frame of the MEN technique. To validate the theoretical formulation, several shielded filtering microstrip structures are analyzed, showing good agreement with respect to both other commercial tools and experimental measurements.
\end{abstract}


\IEEEpeerreviewmaketitle

\section{Introduction}
\IEEEPARstart{C}{urrently},  computational electromagnetics techniques \cite{libro_waveguide} are widely used to save development time and manufacturing costs in the microwave industry. Consequently, new more efficient numerical methods for designing microwave components are of great interest in the electromagnetics research area. What is needed are techniques that can perform efficient and accurate analysis that provide results very close to the experimental data. A large variety of numerical techniques have indeed been reported in the technical literature \cite{libro_waveguide}, and are also implemented in commercial software tools such as ANSYS HFSS, CST Microwace Studo, FEST3D or ADS \cite{FEM, FDTD,IE}. Some of these techniques  \cite{FEM, FDTD} are considered generic, since they can perform the analysis of arbitrary microwave structures, at the expense, most of the time, of consuming high computational resources.  On the other hand, more specific techniques, as those based on modal methods and integral equations \cite{IE}, are able to carry out the analysis of a more limited variety of geometries, but with important reductions in computational time.

In this context, the MEN formulation \cite{libro_marco} has proved to be an efficient and accurate tool for the analysis of waveguide components. This technique is based first on the separate analysis of the discontinuities that are part of the structure under study, and then on their combination to form a computationally efficient multimode equivalent network to represent the complete device. For each discontinuity, the formulation starts by imposing the corresponding boundary conditions to obtain relevant integral equations. A circuit interpretation of the integral equations leads to an equivalent network, where higher order modes that are excited on both sides of the discontinuity are rigorously accounted for with a generalized impedance or admittance coupling matrix.

The MEN technique has been traditionally employed for the analysis of waveguide components, whose usefulness and efficiency are shown in \cite{art:2D}. However, microstrip circuits have become a consolidated technology very popular in the microwave industry. The main goal of this work is to extend the MEN technique to the analysis of practical  multilayer planar circuits. The first contribution in the context of the MEN formulation can be found in \cite{Electric_zeroth_MEN}, where the authors derived the electric and magnetic MEN formulations applied to the analysis of zero-thickness discontinuities (metal-strip gratings seen as apertures or obstacles). However, these original formulations do not include the ports in the transverse plane that are needed for the analysis of  microstrip circuits. The first works that can be found in this direction are reported in \cite{paper_mosig}, \cite{art:miNEMO2018}. In this last work, the authors derived the MEN formulation for the analysis of multilayer planar circuits composed of rectangular printed metallizations using an electric field formulation. Still, the formulation in \cite{art:miNEMO2018} is limited to areas with a rectangular shape, where entire domain basis functions are known analytically.

In this paper, we present an extension of the method proposed in \cite{art:miNEMO2018} to the case of metallic patches with arbitrary shape (like the one shown in Fig. \ref{fig:step}). Another difference with respect to \cite{art:miNEMO2018} is that a magnetic formulation \cite{balanis} is followed to derive the corresponding integral equations. To obtain the basis functions needed to solve the integral equations with complex planar geometries, the BI-RME (Boundary Integral-Resonant Mode Expansion)  method \cite{paper_original_birme, paper_birme_stephan} will be employed.

The extension proposed in this paper is also related to the work published in \cite{paper_birme}. In that work, the BI-RME method was also used  for the analysis of planar structures of complex geometries, but using an electric field integral equation approach.  The main differences with respect to this previous contribution are that now, the problem is solved in the common frame of the MEN representation, and that the formulation is based on the magnetic field boundary condition (that relies on the aperture and not on the metal parts).  In comparison to other integral equation techniques \cite{CAD}, the most important advantage of this method is related to the analysis of structures containing several levels of metallizations. In the general integral equation technique, all the unknowns of the problem are included in the same integral equation. This can lead to matrices of very large sizes. On the contrary, the MEN technique formulates one separate integral equation for each discontinuity, whose solution provides an impedance or admittance matrix representation. Then, all discontinuities are coupled together by cascading the different impedance/admittance multimode coupling matrices. This leads to a reduction in the complexity of the integral equations that must be solved to accomplish the rigorous full-wave analysis of multilayer structures.

The manuscript is organized as follows. In Section \ref{secc:formulation} we describe the theoretical derivations.  To validate the theory, several numerical results of practical filtering microstrip structures are then discussed in Section \ref{secc:results}. Finally, the main conclusions of this contribution are outlined in Section \ref{secc:conclusiones}.

\section{Formulation} \label{secc:formulation}
\begin{figure}[!h]
	\centering
	\includegraphics[scale=0.68]{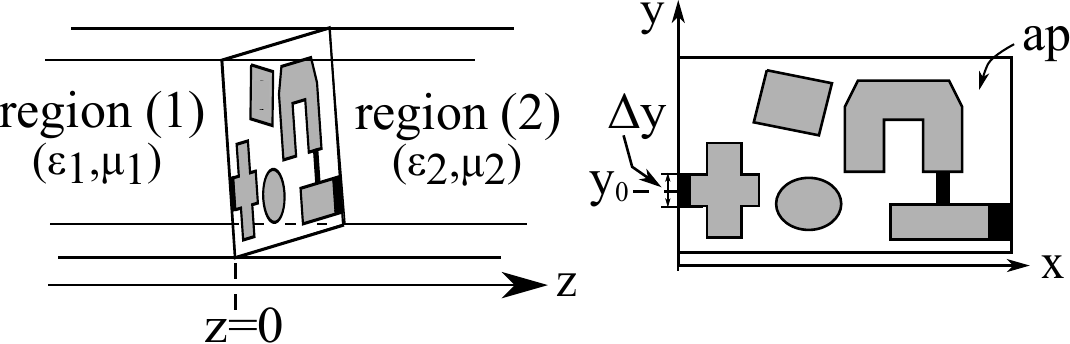}
	\vspace{-0.09in}
	\caption{Two dimensional zero-thickness printed circuit in planar technology. The gray areas correspond to the printed metallizations and the dark areas to the excitation and internal ports, both placed in the transverse (z=0) plane. The media on both sides of the discontinuity can be different, for instance the substrate and the air of a microstrip structure.}
	\label{fig:step}
\end{figure}

In this section, we formulate the rigorous MEN representation of a two dimensional zero-thickness obstacle consisting of  arbitrarily shaped metallizations (see Fig.~\ref{fig:step}). The theoretical derivations also include internal and external lumped ports in the transverse plane, as shown in Fig.~\ref{fig:step}. An important difference with respect to \cite{art:miNEMO2018} is that the magnetic field boundary condition is used to derive the new formulation.

For the sake of simplicity just one lumped excitation port is considered during the discussion of the theory. The extension to more than one port can be easily accomplished following the same procedure. Furthermore, the circuit can also be excited using standard waveguide modes, as it is done when using the MEN for the analysis of waveguide devices.

The formulation starts by imposing the continuity of the tangential component of the magnetic field $\mathbf{H}_{t}^{(\delta)}$ in the aperture at $z=0$, including the excitation contribution
\begin{equation}
\mathbf{H}_{t}^{(1)} (s)- \mathbf{H}_{t}^{(2)} (s) =  \mathbf{z}_{0} \times \mathbf{J}_{exc}  \label{eq:BC}
\end{equation}
where $\delta=1$ or $\delta=2$ for $z\leq 0$ or $z\geq 0$, respectively, and  $s$ indicates a point in the junction cross-section of Fig.~\ref{fig:step}. In eq. (\ref{eq:BC}), the excitation port has been accounted for by the surface electric current density $\mathbf{J}_{exc}$. 
In this paper, we use a pulse excitation model, which can be considered as a more elaborated version of the delta-gap excitation model \cite{paper_delta}. In this model, a current $I_{0}$ constant along the width of a pulse function is assumed to be applied at the exciting port. Therefore, the expression for the surface electric current density $\mathbf{J}_{exc} $ is the following
\begin{equation}
	\mathbf{J}_{exc}=-I_0 \, \frac{1}{\Delta y} \prod \left(\frac{y-y_0}{\Delta y}\right) \mathbf{x}_{0}  \label{eq:exc_pulse}
\end{equation}
where $y_0$ is the y-axis coordinate of the center of the pulse and $\Delta y$ is the width (see Fig. 1). The amplitude of the pulse is normalized with the  $1/\Delta y$ factor to ensure that the power flow $P_{exc}$ defined as
\begin{equation}
P_{exc}= \frac{1}{2} \int_{S_{exc}} \left( \mathbf{E}_{exc} \times \mathbf{H}_{exc}^{*} \right) \, \mathbf{z}_{0} \, ds \label{eq:p_exc}
\end{equation}
 is equal to that associated to the equivalent transmission line $P_{0}$ defined as
\begin{equation}
P_{0}= \frac{1}{2} V_0 {{I}_{0}}^{*} \label{eq:p_0}
\end{equation}
where $V_0$ is the voltage across the port. Moreover, the integral in eq. (\ref{eq:p_exc}) extends to the pulse area denoted as ($S_{exc}$).

By using this excitation model, and the well known modal expansion formalism \cite{libro_marco}, eq. (\ref{eq:BC}) can be written as
\begin{equation}
	\sum_{m=1}^{+\infty} I_{m}^{(1)} \mathbf{h}_{m}^{(1)} (s) -\sum_{m=1}^{+\infty} I_{m}^{(2)} \mathbf{h}_{m}^{(2)} (s)=   \frac{-I_0 }{\Delta y} \prod \left(\frac{y-y_0}{\Delta y}\right) \mathbf{y}_{0} \label{eq:mag}
\end{equation}
where $\mathbf{h}_{m}^{(\delta)} (s)$ represents the magnetic vector mode function of mode $m$ in the medium $(\delta)$, and $I_{m}^{(\delta)}$ refers to the modal current. It is important to note that $m$ refers to \textit{all TE and TM modes}. To continue, we separate the accessible from localized modes by splitting the infinite series as
\begin{equation}
\begin{split}
&\sum_{n=1}^{N(1)} I_{n}^{(1)} \mathbf{h}_{n}^{(1)} (s) -\sum_{n=1}^{N(2)} I_{n}^{(2)} \mathbf{h}_{n}^{(2)} (s)+ \frac{ I_0}{\Delta y} \prod \left(\frac{y-y_0}{\Delta y}\right) \mathbf{y}_{0} \\
&= \sum_{m=N(1)+1}^{+\infty}  \frac{V_{m}^{(1)}}{Z_{m}^{(1)}} \mathbf{h}_{m}^{(1)} (s)+\sum_{m=N(2)+1}^{+\infty}  \frac{V_{m}^{(2)}}{Z_{m}^{(2)}}  \mathbf{h}_{m}^{(2)} (s) \label{eq:acc_loc}
\end{split}
\end{equation}
where $N(\delta)$ is the number of accessible modes in each region. In eq. (\ref{eq:acc_loc}), $V_{m}^{(\delta)}$ and $Z_{m}^{(\delta)}$ represent the total modal voltage and the characteristic modal impedance, respectively. Note that, for the sake of clarity, the subscript on the left hand side in eq. (\ref{eq:acc_loc}) has been changed  to $n$.

We now recognize that, since the metallic circuit is zero-thickness, and the waveguide cross section in region (1) is identical to the one in region (2), the magnetic vector mode functions in region (1) $\big(\mathbf{h}_{m}^{(1)}\big)$ are identical to the ones in region (2) $\big(\mathbf{h}_{m}^{(2)}\big)$. Then, knowing that $\mathbf{h}_{m}^{(1)} (s)=\mathbf{h}_{m}^{(2)} (s)=\mathbf{h}_{m} (s)$ and ${V}_{m}^{(1)} ={V}_{m}^{(2)} ={V}_{m} $, eq. (\ref{eq:acc_loc}) is transformed into
\begin{equation}
\begin{split}
\sum_{n=1}^{N} \overline{I}_{m} \mathbf{h}_{n} (s) + & I_0 \, \frac{1}{\Delta y} \prod \left(\frac{y-y_0}{\Delta y}\right) \mathbf{y}_{0}=\\
& \sum_{m=N+1}^{+\infty}  V_{m} Y_{m}^{T} \mathbf{h}_{m} (s) \label{eq:acc_zth}
\end{split}
\end{equation}

In eq. (\ref{eq:acc_zth}), $\overline{I}_{m}$ represents the total modal current at the interface, and is defined as $\overline{I}_{m}= I_{m}^{(1)}- I_{m}^{(2)}$. Also, $Y_{m}^{T}$  is the  total characteristic modal admittance defined as $Y_{m}^{T}= Y_{m}^{(1)} + Y_{m}^{(2)}$, where $Y_{m}^{(\delta)}$ represents the characteristic modal admittance of the mode $m$ in the medium $(\delta)$. Note that the number of accessible modes in both regions is assumed to be equal, $N(1)=N(2)=N$. A different number of accessible modes can also be easily analyzed.

To continue the formulation, we now recall the following relation between the modal voltage ${V}_{m}$ and the electric field in the aperture $\mathbf{E}$ 
\begin{equation}
	{V}_{m} = \int_{ap} \left[\mathbf{z}_{0} \times \mathbf{E} \right] \cdot \mathbf{h}_{m}^{*} (s')\, ds' \label{eq:voltage_modal}
\end{equation}
where the integral extends over the entire aperture ($ap$) at the junction (see Fig. \ref{fig:step}).
 
Due to the linearity of the problem, the unknown electric field in the aperture $(\mathbf{z}_{0} \times \mathbf{E})$ must be linearly proportional to the excitations, so that it can be expanded into a sum of partial surface magnetic current densities $\mathbf{M}_{n} (s') $ and $\mathbf{M}_{0} (s') $ as follows
\begin{equation}
\left[\mathbf{z}_{0} \times \mathbf{E} \right] = \sum_{n=1}^{N}  \overline{I}_{m}\, \mathbf{M}_{n} (s') + I_0 \,\mathbf{M}_{0} (s') \label{eq:campo_apertura}
\end{equation}
where $ \mathbf{M}_{n}$ and $\mathbf{M}_{0}$ are as yet the unknown functions of our problem.
Substituting now eq. (\ref{eq:campo_apertura}) into eq. (\ref{eq:voltage_modal}), and by using the resulting equation in eq. (\ref{eq:acc_zth}), we can finally obtain the fundamental integral equations of our problem
\begin{equation}
\begin{split}
\mathbf{h}_{n} (s)=  \int_{ap}  \mathbf{M}_{n} (s') \, \mathbf{K} (s,s') ds'
\label{eq:eqI1}
\end{split}
\end{equation}

\begin{equation}
\begin{split}
\frac{1}{\Delta y} \prod \left(\frac{y-y_0}{\Delta y}\right) \mathbf{y}_{0}=  \int_{ap}  \mathbf{M}_{0} (s') \, \mathbf{K} (s,s') ds'
\label{eq:eqI2}
\end{split}
\end{equation}
where $\mathbf{K} (s,s')$ is the kernel of the integral equation 
\begin{equation}
\mathbf{K} (s,s') =  \sum_{m=N+1}^{+\infty} Y_{m}^{T} \mathbf{h}_{m}^{*} (s') \, \mathbf{h}_{m} (s) \label{eq:kernel}
\end{equation}
To complete the rigorous network formulation, eq. (\ref{eq:campo_apertura}) is used in eq. (\ref{eq:voltage_modal}) to write the modal voltages in terms of the modal currents

\begin{equation}
	V_{0} =  Z_{0,0}\,I_{0}   + \sum_{n=1}^{N} Z_{0,n}\,\overline{I}_{n}    \label{eq:voltage_modal_final_2}
\end{equation}
\begin{equation}
	V_{m} = Z_{m,0}\,I_{0}  + \sum_{n=1}^{N} Z_{m,n}\, \overline{I}_{n}    \label{eq:voltage_modal_final_1}
\end{equation}
where $ Z_{0,0}, \, Z_{0,n},  \, Z_{m,0}  $ and $Z_{m,n}$ are the elements of the generalized impedance coupling matrix $\mathbf{\overline{Z}}$
\begin{equation}
	Z_{0,0}  =  \int_{ap}  \mathbf{M}_{0} (s') \cdot \left(  \frac{1}{\Delta y} \prod \left(\frac{y'-y_0}{\Delta y}\right) \mathbf{y}_{0} \right) ds' \label{eq:Z00}
\end{equation}
\begin{equation}
	Z_{0,n}  =  \int_{ap}  \mathbf{M}_{n} (s') \cdot \left(  \frac{1}{\Delta y} \prod \left(\frac{y'-y_0}{\Delta y}\right) \mathbf{y}_{0} \right) ds' \label{eq:Z0n}
\end{equation}
\begin{equation}
	Z_{m,0}  =  \int_{ap}  \mathbf{M}_{0} (s') \cdot \mathbf{h}_{m}^{*} (s') \, ds' \label{eq:Zm0}
\end{equation}
\begin{equation}
	Z_{m,n}  = \int_{ap}  \mathbf{M}_{n} (s') \cdot \mathbf{h}_{m}^{*} (s')  \, ds' \label{eq:Zmn}
\end{equation}

\begin{figure}[!h]
	\centering
	\includegraphics[scale=0.85]{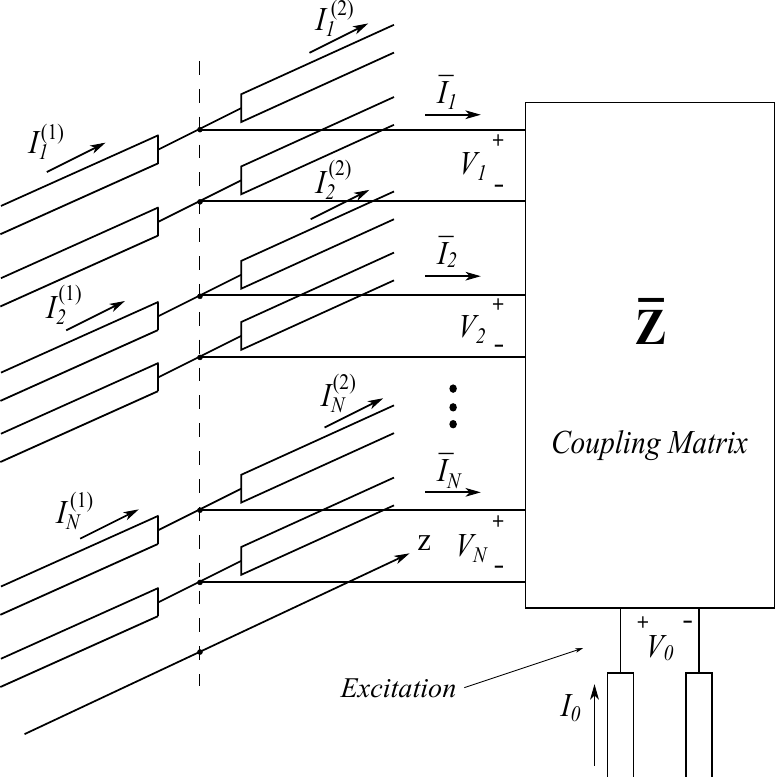}
	\vspace{-0.09in}
	\caption{Multimode Equivalent Network (generalized Z-matrix) representation for the zero-thickness discontinuity with arbitrary shape metallic areas depicted in Fig. \ref{fig:step}, including a port in the transverse plane.}
	\label{fig:MEN}
\end{figure}

With these elements, the multimode impedance coupling matrix is defined as follows
\begin{equation}
\mathbf{\overline{Z}}=
\left[ {\begin{array}{ccccc}
	Z_{0,0} & Z_{0,1} & Z_{0,2}& ... & Z_{0,N}\\
	Z_{1,0} & Z_{1,1} & Z_{1,2}& ... & Z_{1,N}\\
	. & . & \,& \, & .\\
	. & \, &. & \, & .\\
	. & \, &\, & . & .\\
	Z_{M,0} & Z_{M,1} & Z_{M,2}& ... & Z_{M,N}\\	
	\end{array} } \right]
\end{equation}
and the network representation is displayed in the form shown in Fig. \ref{fig:MEN}.

In order to solve the integral equations in eq. (\ref{eq:eqI1}) and eq. (\ref{eq:eqI2}), we now expand the unknown partial magnetic current functions $\mathbf{M}_{n}(s')$ and $\mathbf{M}_{0}(s')$  using  the Method of Moments (MoM) \cite{libro_marco}. In particular, the magnetic vector mode functions in the air (aperture), at the junction plane, are used here as basis ($\mathbf{h}_{k}$) and test ($\mathbf{h}_{i}$) functions, applying Galerking procedure \cite{libro_marco}. Following this approach, the unknowns $\mathbf{M}_{n} (s')$ and $\mathbf{M}_{0} (s')$ can finally be written as
\begin{equation}
	\mathbf{M}_{0} (s')  =  \sum_{k=1}^{N_b} \alpha_{0,k} \, \mathbf{h}_{k}(s') \, ; \, \mathbf{M}_{n} (s')  =  \sum_{k=1}^{N_b} \alpha_{n,k} \, \mathbf{h}_{k} (s')   \label{eq:MoM2}
\end{equation}
where $N_b$ is the numerical parameter that determines the number of basis and test functions used in the MoM procedure, and $\alpha_{0,k}$  and $\alpha_{n,k}$ are the unknown constant coefficients of the expansion. 

Applying the test procedure, the integral equations in eq. (\ref{eq:eqI1}) and eq. (\ref{eq:eqI2}) are transformed into the following system of linear equations
\begin{equation}
	\begin{split}
		C_{i,n}=  \sum_{k=1}^{N_b} \alpha_{n,k}  \sum_{m=N+1}^{+\infty} Y_{m}^{T} \, C_{k,m} \, C_{i,m}  
		\label{eq:eqI1c}
	\end{split}
\end{equation}
\begin{equation}
	\begin{split}
		C_{i,0}=  \sum_{k=1}^{N_b} \alpha_{0,k}  \sum_{m=N+1}^{+\infty} Y_{m}^{T}\, C_{k,m}\, C_{i,m}  
		\label{eq:eqI2c}
	\end{split}
\end{equation}

The coefficients $C_{k,m}$ and  $C_{i,m}$ represent the coupling integrals between the localized modes of the rectangular waveguide in regions (1) and (2), and the aperture modes  used as basis and test functions, respectively
\begin{equation}
	C_{k,m}  =  \int_{ap}  \mathbf{h}_{k} (s') \cdot \mathbf{h}_{m}^{*} (s') \, ds' \label{eq:Ckm}
\end{equation}
\begin{equation}
	C_{i,m}  =  \int_{ap}  \mathbf{h}_{i} (s) \cdot \mathbf{h}_{m} (s)  \,ds \label{eq:Cim}
\end{equation}

The terms $C_{i,n}$  are the coupling integrals between the accessible modes of the rectangular waveguide in regions (1) and (2) and the aperture modes used as test functions
\begin{equation}
	C_{i,n}  =  \int_{ap}  \mathbf{h}_{i} (s) \cdot \mathbf{h}_{n} (s)  ds \label{eq:Cin}
\end{equation}

The term $C_{i,0}$, on the other hand, refers to the coupling integral between the test functions and the excitation pulse
\begin{equation}
	C_{i,0}  =  \int_{ap}  \mathbf{h}_{i} (s) \cdot \left( \frac{1}{\Delta y} \prod \left(\frac{y-y_0}{\Delta y}\right) \mathbf{y}_{0} \right) ds \label{eq:Ci0}
\end{equation}

It is important to mention that, since the metallizations at the junction are arbitrarily shaped, the coupling integrals in eq. (\ref{eq:Ckm}), eq. (\ref{eq:Cim}) and eq. (\ref{eq:Cin}), that are related to the air modes in the aperture, can be conveniently computed numerically using the BI-RME method \cite{paper_original_birme, paper_birme_stephan}. The coupling integrals $C_{i,0}$ can then be computed analytically using the numerical expansion in terms of rectangular waveguide modes of the modes in the aperture, which are also provided by the BI-RME method. In particular, let  $h_m(s)$ be the rectangular waveguide modes used to expand the arbitrary aperture modes $h_i (s)$. Using the BI-RME method we obtain the following numerical expansion
\begin{equation}
\mathbf{h}_{i} (s)  =  \sum_{m=1}^{N_k} C_{i,m} \, \mathbf{h}_m(s) \label{birme_exp}
\end{equation}
where $C_{i,m} $ are the expansion coefficients calculated by BI-RME (the same ones as those in eq. (\ref{eq:Cim})). According to eq. (\ref{birme_exp}), the coupling integral in eq. (\ref{eq:Ci0}) can be computed as
\begin{equation}
C_{i,0}  =  \sum_{m=1}^{N_k} C_{i,m}  \int_{ap}  \mathbf{h}_{m} (s) \cdot \left( \frac{1}{\Delta y} \prod \left(\frac{y-y_0}{\Delta y}\right) \mathbf{y}_{0} \right) ds \label{eq:Ci0_f}
\end{equation}
where the integral between the rectangular waveguide modes and the pulse function is now analytical.

According to this notation, the elements of the generalized impedance coupling matrix $\mathbf{\overline{Z}}$ can be rewritten in a more compact form as
\begin{equation}
	Z_{0,0}  =  \sum_{k=1}^{N_b} \alpha_{0,k} \, C_{k,0}  \quad ; \quad Z_{0,n}  = \sum_{k=1}^{N_b} \alpha_{n,k} \, C_{k,0} \label{eq:Z00c}
\end{equation}
\begin{equation}
	Z_{m,0}  =  \sum_{k=1}^{N_b} \alpha_{0,k} \, C_{k,m} \quad ; \quad Z_{m,n}  =  \sum_{k=1}^{N_b} \alpha_{n,k} \, C_{k,m} \quad \label{eq:Zm0c}
\end{equation}

In order to increase the efficiency of the formulation described before, we now use the Kummer's transformation, as already described in \cite{paper_nemo} for the acceleration of the kernel of the integral equations. As shown in \cite{paper_nemo}, using this strategy the kernel is divided into static and dynamic parts. As a result, the dynamic part, that needs to be recomputed per each frequency point, exhibits now a very fast convergence rate.

\section{Numerical Results and Applications} \label{secc:results}
\begin{figure}[!h]
	\centering
	\includegraphics[scale=0.65]{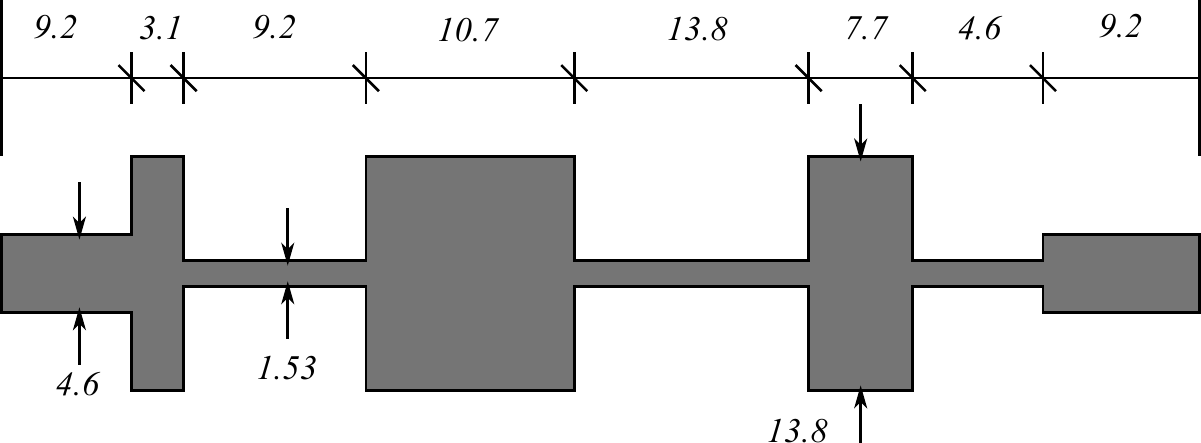}
	\vspace{-0.09in}
	\caption{Low-pass microstrip filter  under study. The dimensions are in mm. The dimensions of the shielding box are: $67.5$ mm $\times\, 67.5$ mm $\times \, 11.4$ mm.  The  dielectric relative permittivity is $\epsilon_{r}=2.33 $ and the substrate thickness is  $t=1.57 $ mm. The filter is centered in the box.}
	\label{fig:low_pass_filter}
\end{figure}

To validate the theoretical derivations, three examples of filtering geometries are analyzed in this section using the novel MEN formulation. The first example consists of a stepped-impedance low-pass filter, first presented in \cite{paper_pasobajo}. The filter structure under analysis is shown in Fig. \ref{fig:low_pass_filter}.

\begin{figure}[!h]
	\centering
	\includegraphics[width=0.43\textwidth]{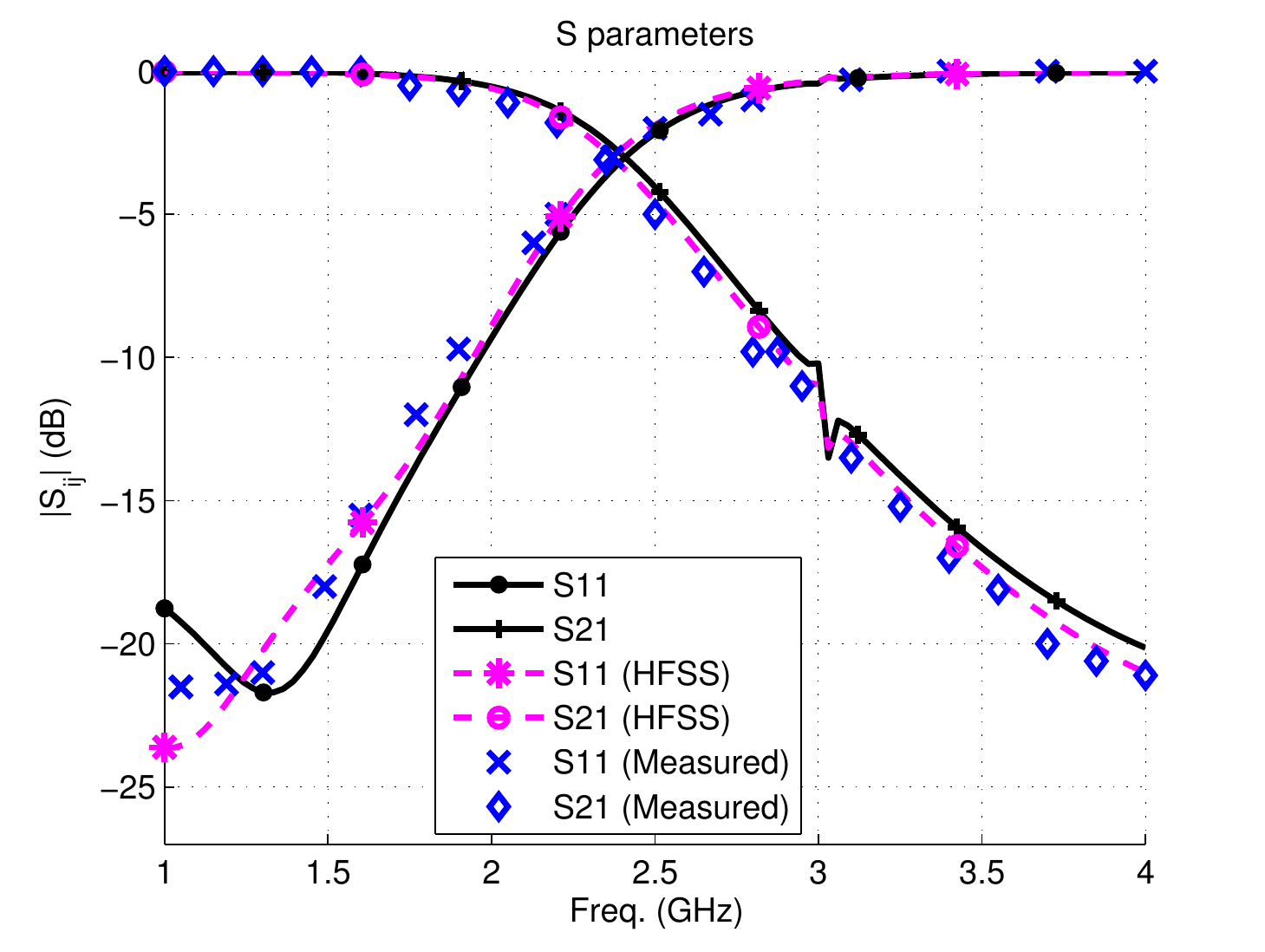}
	\vspace{-0.09in}
	\caption{Scattering parameters of the low-pass filter in Fig. \ref{fig:low_pass_filter} computed using the novel formulation, using a finite elements based electromagnetic analysis tool (HFSS), and compared to the measured response.}
	\label{fig:S_low_pass_filter}
\end{figure}

In Fig. \ref{fig:S_low_pass_filter}, we show the filter response retrieved by using the MEN formulation. In this figure, the results obtained with a commercial software tool (HFSS) and the measured results are also reported. 

As can be noticed, the MEN results show good agreement as compared to both the response provided by HFSS and the measurements. To obtain this result, we have used 4000 terms in the kernel (3900 in the static part and 100 in the dynamic part of the kernel \cite{paper_nemo}). In the MoM and Galerking procedure, 600 basis/test functions (computed using the BI-RME method) have been considered.

From a computational point of view, we first execute BI-RME tool to obtain the relevant numerical coupling integrals, and then we run our MEN code. It is important to consider that, in a frequency sweep analysis, both the static part of the kernel and the coupling integrals are not re-computed for each frequency point, since they are frequency independent. For the results presented above, using an Intel Xeon  E5-2620 v3  @~2.40~GHz with 32 GB of RAM, the computational time required to obtain the coupling integrals is less than 3 minutes. The MEN code takes about 10.4 seconds for all the frequency sweep (100 points).  

\begin{figure}[!h]
	\centering
	\includegraphics[scale=0.65]{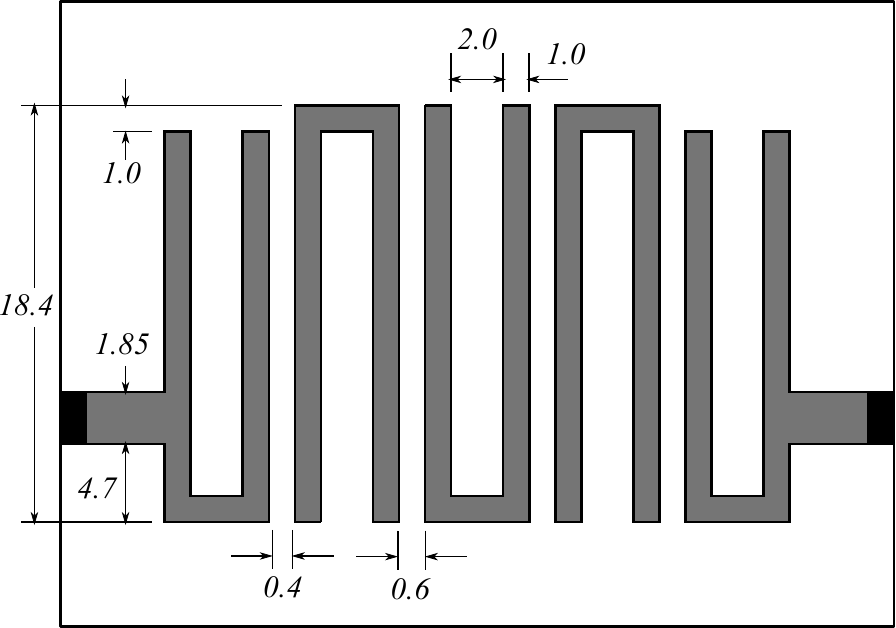}
	\vspace{-0.09in}
	\caption{Hairping band-pass filter geometry under study. The dimensions are in mm.  The dimensions of the shielding box are: $31.2$ mm $\times \, 30$ mm $\times \, 5$ mm. The  dielectric relative permittivity is $\epsilon_{r}=6.15 $ and the substrate thickness is  $t=1.27 $ mm. The filter is centered in the box.}
	\label{fig:hairpin_filter}	
\end{figure}

Another example is presented in Fig. \ref{fig:hairpin_filter}. This is a hairpin band-pass filter with two exciting ports in the transverse plane (this filter was originally proposed in  \cite{libro_hong}).

\begin{figure}[!h]
	\centering
	\includegraphics[width=0.5\textwidth]{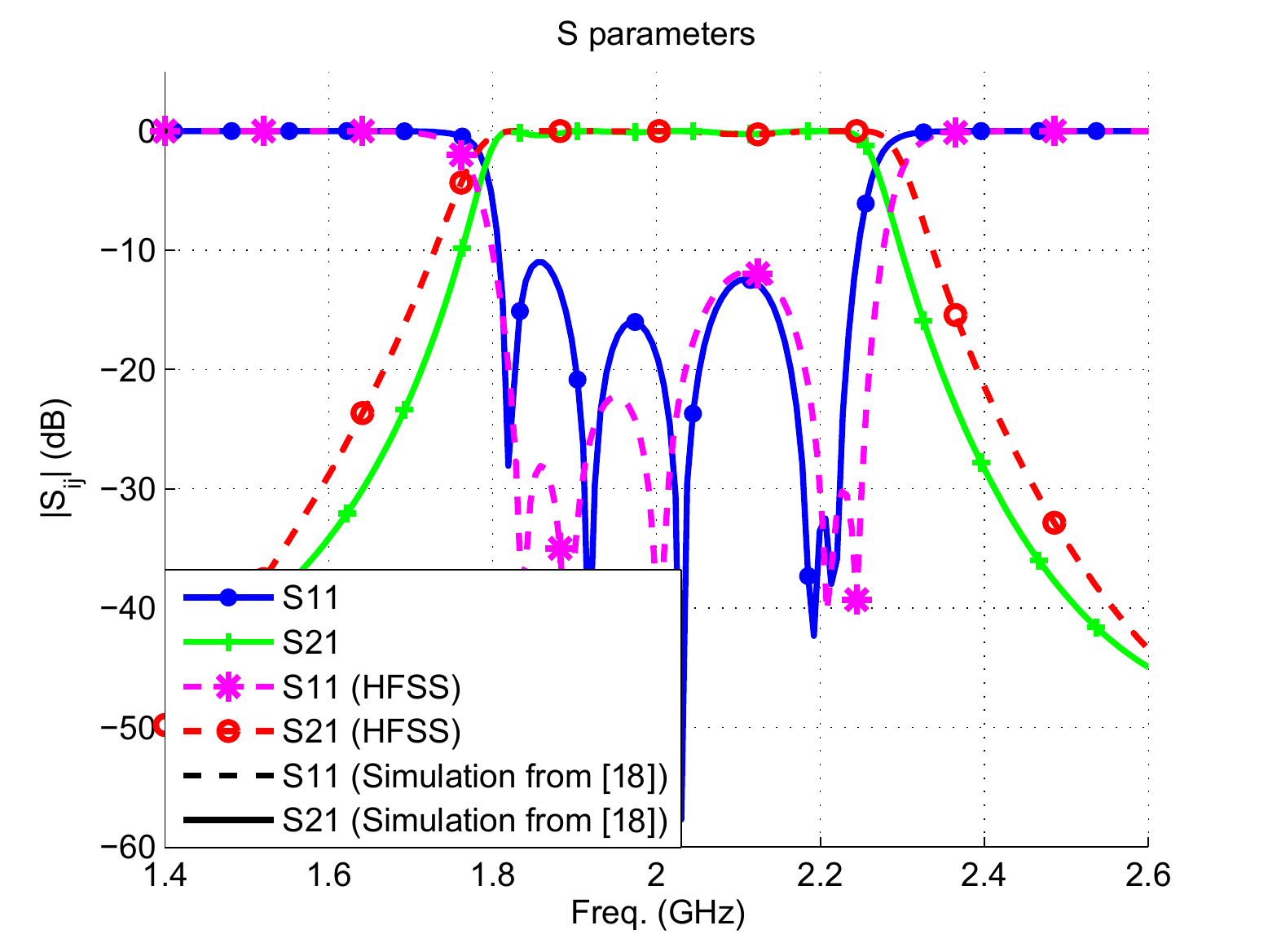}
	\vspace{-0.09in}
	\caption{In-band S-parameters of the hairpin bandpass filter in Fig. \ref{fig:hairpin_filter} using our MEN formulation and both the electromagnetic analysis tool HFSS and the simulation reported in \cite{libro_hong}.}
	\label{fig:S_hairpin_filter_in}
\end{figure}

Fig. \ref{fig:S_hairpin_filter_in} shows the in-band response obtained using our formulation. In this figure, the MEN results are compared to those obtained using the commercial software HFSS and the EM simulated results from  \cite{libro_hong}.

\begin{figure}[!h]
	\centering
	\includegraphics[width=0.43\textwidth]{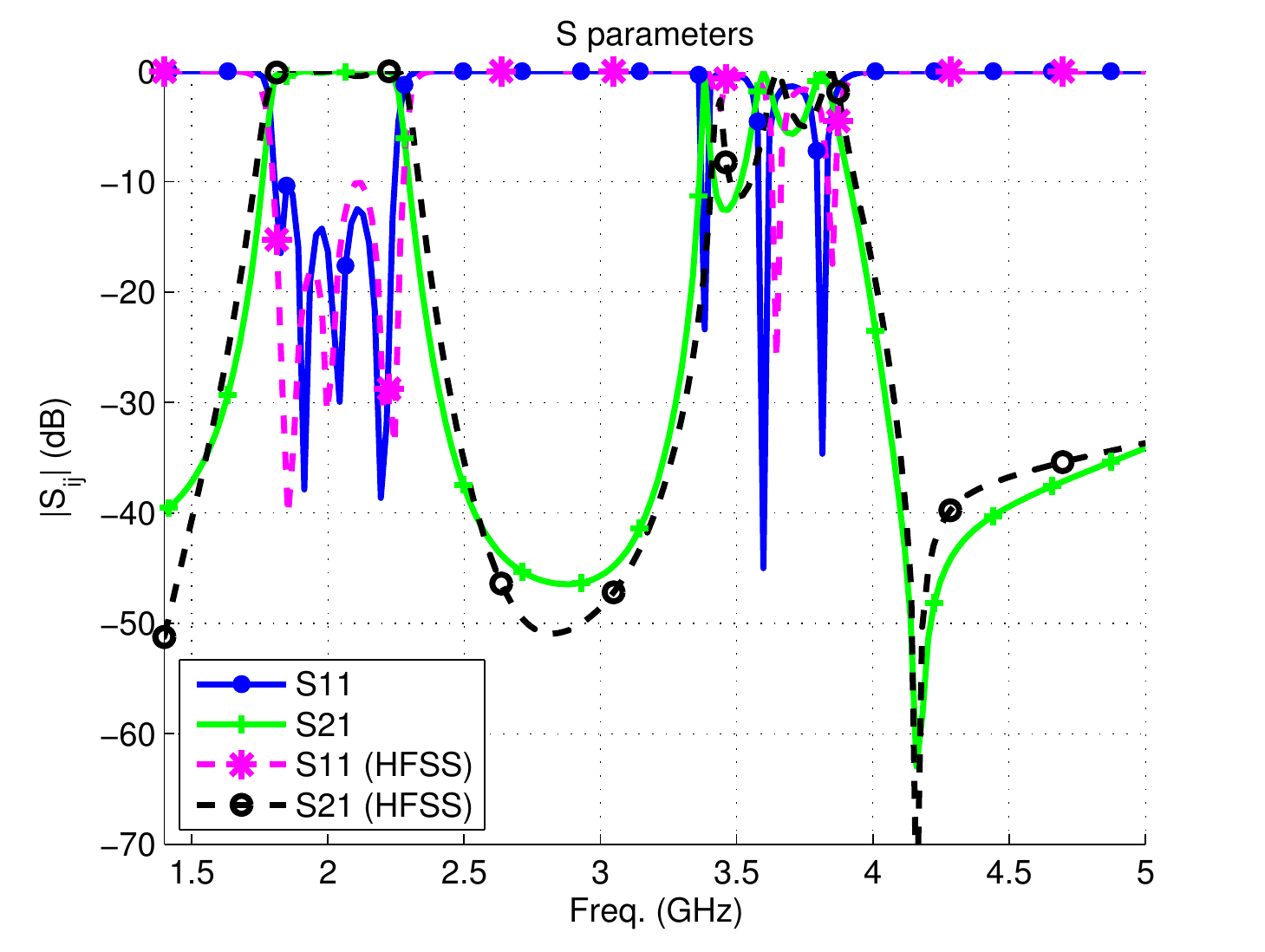}
	\vspace{-0.09in}
	\caption{Out-of-band S-parameters of the hairpin bandpass filter in Fig. \ref{fig:hairpin_filter} using our MEN formulation and the electromagnetic analysis tool HFSS.}
	\label{fig:S_hairpin_filter_out}
\end{figure}

For this example, the numerical parameters needed to obtain the result shown are: 5000 terms in the kernel (4500 in the static part and 500 in the dynamic part of the kernel \cite{paper_nemo}) and 310 basis/test functions in the MoM procedure. The computational time required for the MEN simulation is about 5.9 seconds (200 frequency points). Additionally, Fig. \ref{fig:S_hairpin_filter_out} shows the out-of-band response of this filter.

As can be seen, in both cases (in-band and out-of-band) the MEN results show good agreement with respect to HFSS results and the simulation data  reported in \cite{libro_hong}.

\begin{figure}[!h]
	\centering
	\includegraphics[scale=0.65]{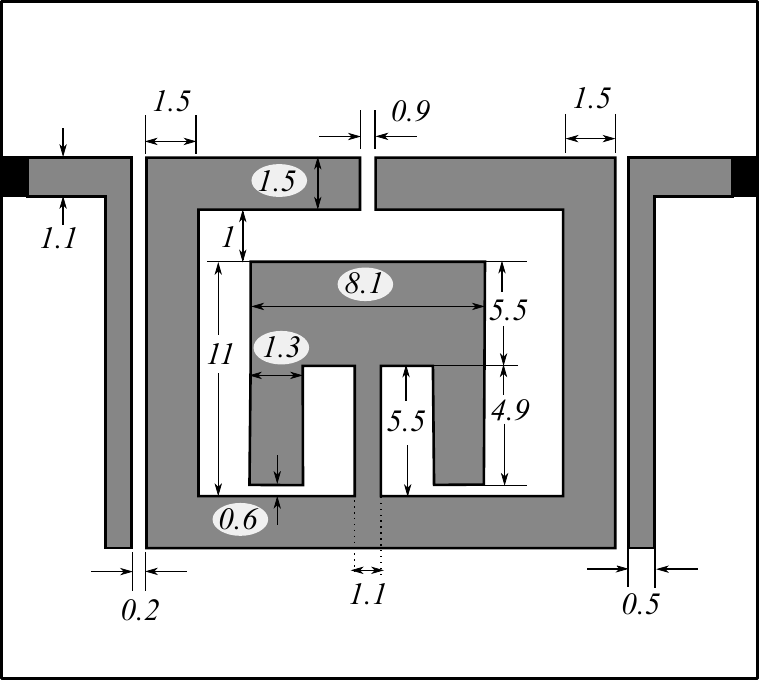}
	\vspace{-0.09in}
	\caption{Dual-mode microstrip band-pass filter  under study (firstly presented in \cite{paper_dualmode_openloop}). The dimensions are in mm.  The dimensions of the shielding box are: $25$ mm $\times \, 20$ mm $\times \, 10$ mm. The  dielectric relative permittivity is $\epsilon_{r}=10.8 $ and the substrate thickness is  $t=1.27 $ mm. The filter is centered in the box.}
	\label{fig:dual_mode}	
\end{figure}

The third example refers to the analysis of a dual-mode microstrip filter, first presented in \cite{paper_dualmode_openloop}. The geometry of the microstrip circuit with the excitation port locations is shown in Fig. \ref{fig:dual_mode}.

\begin{figure}[!h]
	\centering
	\includegraphics[width=0.43\textwidth]{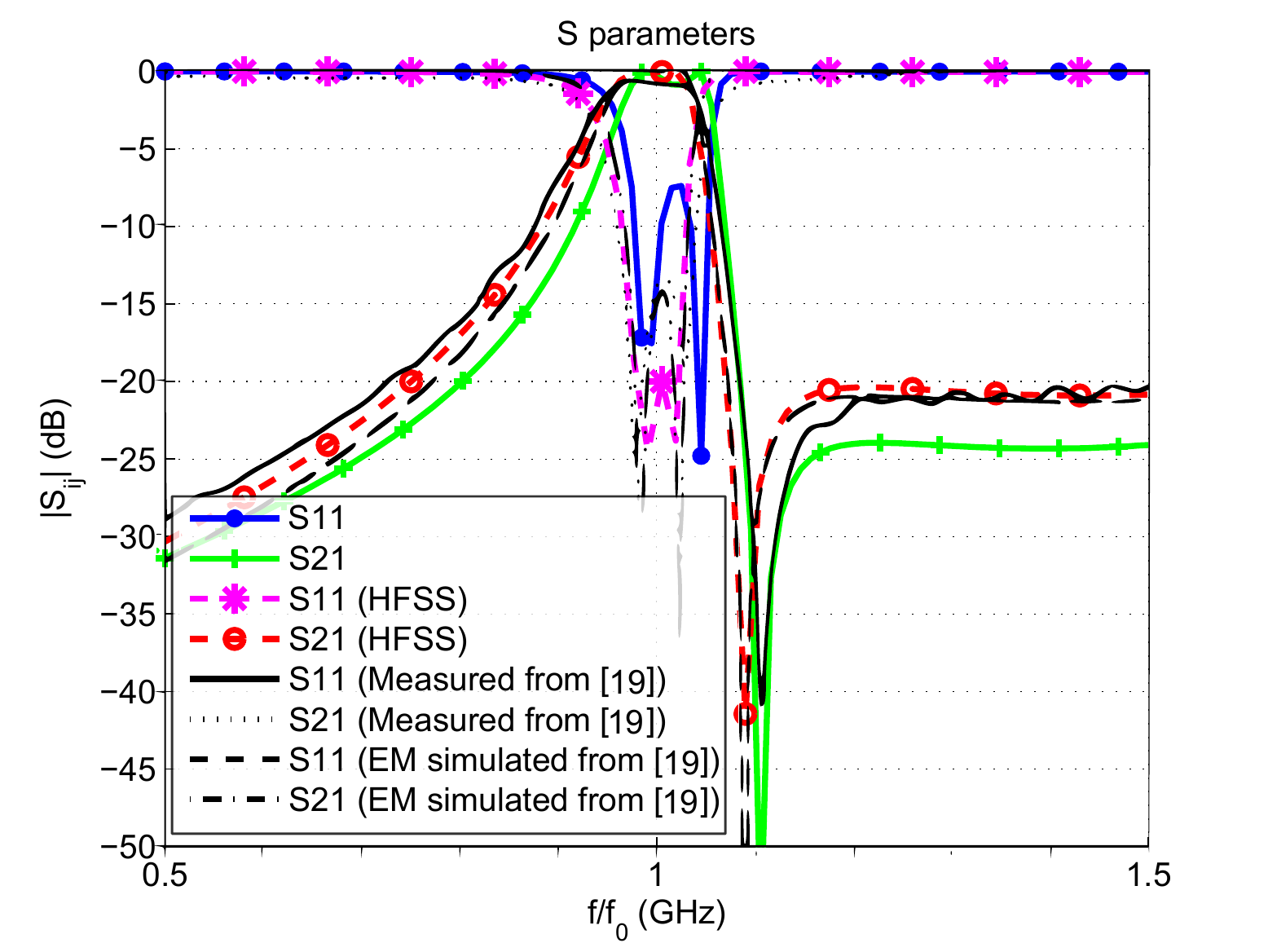}
	\vspace{-0.09in}
	\caption{In-band S-parameters of the dual-mode bandpass filter in Fig. \ref{fig:dual_mode} using our MEN formulation, the electromagnetic analysis tools HFSS, the results of EM simulations (using Sonnet), and measurements from \cite{paper_dualmode_openloop}.}
	\label{fig:s_dual_mode_inband_d1}
\end{figure}

In Fig. \ref{fig:s_dual_mode_inband_d1}, we show the passband response obtained using our formulation. This response is compared to the HFSS response and also to the EM simulations and measurements reported in \cite{paper_dualmode_openloop}. The EM simulations reported in \cite{paper_dualmode_openloop} are obtained using the commercially available full-wave electromagnetic simulator Sonnet. As it is reported in \cite{paper_dualmode_openloop}, all the simulations are normalized in the frequency axis using $f_0=1.07$ GHz (design center frequency) and the measurements are normalized using $f_0=0.9975$ GHz (measured center frequency).

To obtain the in-band response shown, 3500 terms have been employed in the kernel (3200 in the static part and 300 in the dynamic part of the kernel \cite{paper_nemo}). In the MoM and Galerking procedure, 300 basis/test functions (computed using the BI-RME method) have been considered. The computational time required to obtain the coupling integrals is less than 7 minutes. The MEN code takes about 2.68 seconds for the complete frequency sweep (100 points).

\begin{figure}[!h]
	\centering
	\includegraphics[width=0.43\textwidth]{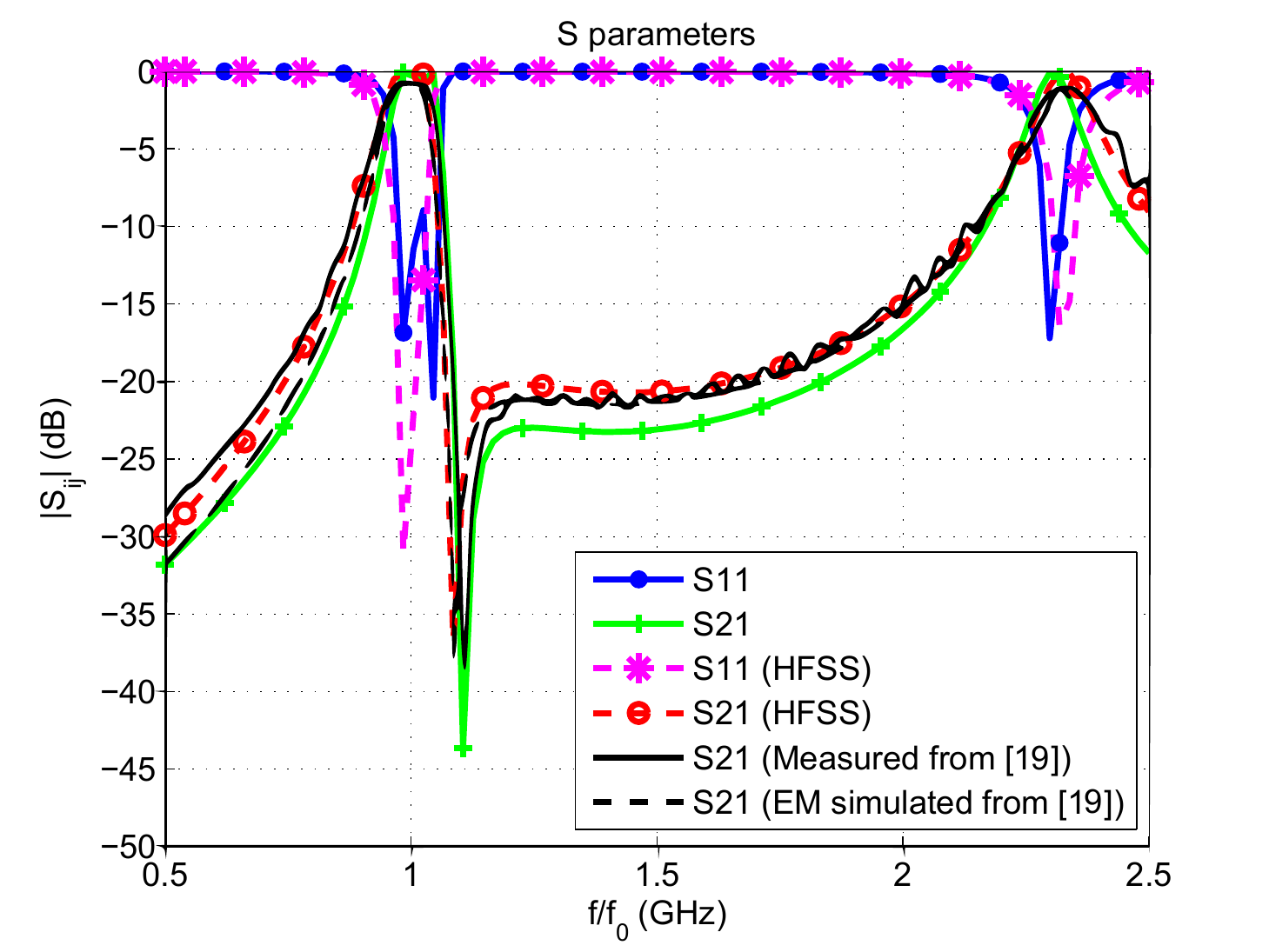}
	\vspace{-0.09in}
	\caption{Out-of-band S-parameters of the dual-mode bandpass filter in Fig. \ref{fig:dual_mode} using our MEN formulation, the electromagnetic analysis tools HFSS, the results of EM simulations (using Sonnet), and measurements from \cite{paper_dualmode_openloop}.}
	\label{fig:S_dual_mode_outband}
\end{figure}

The out-of-band response is shown in Fig. \ref{fig:S_dual_mode_outband}, where the results obtained using the MEN technique are also compared to HFSS response and the data from \cite{paper_dualmode_openloop}.
In this example, the results are also in good agreement with the theoretical and experimental data taken from \cite{paper_dualmode_openloop}. 

One very important point that must be noted is that, for the three examples discussed, just one accessible mode is needed, since the structures contain only one discontinuity (one metallization layer). 

It is also important to mention that using in HFSS a selectively refined mesh around the critical gaps of the structures analyzed, in average, more than 25 minutes (for 100 frequency points) are needed in order to obtain the corresponding presented results.

With the examples presented, the theory for the rigorous analysis of zero-thickness printed circuits composed of arbitrarily shaped metallizations has been fully validated. Using this new MEN formulation any arbitrary shape metallic areas can be included in the structure. The results obtained clearly demonstrate that the analysis of practical shielded printed circuits, using the  MEN extension proposed in this contribution, is both efficient and accurate.

\section{Conclusion} \label{secc:conclusiones}
In this paper, the rigorous MEN technique has been extended to the analysis of planar discontinuities that include arbitrarily shaped metallizations together with internal and external ports in the transverse plane. Using the reported novel formulation, a very large variety of practical  boxed microwave printed components can be efficiently analyzed in the common frame of the MEN technique. It is important to note that this contribution enables the rigorous and efficient analysis of integrated waveguide and planar circuits.  In addition to theory, the analysis of a number of filtering microstrip structures has been presented,  showing good agreement with respect to other electromagnetic analysis tools and with real measurements. The examples presented fully validate the proposed theoretical formulation, and clearly confirm the usefulness and efficiency of the novel procedure in the analysis of real planar microstrip devices.

\ifCLASSOPTIONcaptionsoff
  \newpage
\fi

\end{document}